\documentclass[a4paper,twocolumn]{esapub} 
\usepackage{graphicx}
\title{A BeppoSAX-WFC viewpoint of new INTEGRAL sources, particularly~IGR~J17544-2619}
\author{Jean in 't Zand}
\author{John Heise}
\affil{SRON National Institute for Space Research \&
Astronomical Institute of Utrecht University, Netherlands}
\author{Pietro Ubertini}
\author{Angela Bazzano}
\affil{CNR Istituto Astrofisica Spaziale e Fisica Cosmica, Sezione Roma, Italy}
\author{Craig Markwardt}
\affil{NASA Goddard Space Flight Center \& University of Maryland, U.S.A.}

\def\ecs{erg~cm$^{-2}$s$^{-1}$}

\def\bron{IGR~J17544-2619}
\def\fdg{\hbox{$.\!\!^\circ$}}
\def\fs{\hbox{$.\!\!^{\rm s}$}}
\def\farcm{\hbox{$.\mkern-4mu^\prime$}}

\begin{document}


\maketitle

\begin{abstract}
Of the 21 new sources that INTEGRAL discovered up to Feb. 1, 2004, 5
were detected with the BeppoSAX Wide Field Cameras at earlier
times. IGR~J16320-4751 appears to be a persistently active X-ray
source which hints at a supergiant Roche-lobe overflowing companion
star in this proposed high-mass X-ray binary.  IGR~J17091-3624 is a
transient source that was detected in 1996 and 2001 with a maximum
flux of 20 mCrab (2-28 keV). It is either a Be X-ray binary or a low
mass X-ray binary transient. IGR~J18483-0311 may be a high-mass X-ray
binary, because it is located in a region rich of such objects, just
like IGR~J19140+098. IGR~J17544-2619 appears to be a frequently active
X-ray source whose hours-long flares, of which WFC detected five, are
reminiscent of the stellar black hole source V4641~Sgr. We discuss
this source in detail.

\end{abstract}

\section{Introduction}

Up to Feb. 1, 2004, 21 new sources were detected with INTEGRAL
according to reports in the literature, see Table~\ref{tab1}. They are
all within 7 degrees from the Galactic plane and it is likely that
most are located in the Galaxy.  Some are obscured in the classical
2-10 keV band by thick absorption columns ($N_{\rm
H}>10^{23}$~cm$^{-2}$), others are quick transients with activity
periods of less than a day. Both kinds are interesting. If they are
X-ray binaries, this suggests deviations from pure Roche-lobe overflow
which is the standard in low-mass X-ray binaries. The thick local
absorption columns could be indicative of massive in or outflows
outside the accretion disk plane (if eclipses nor dips are present),
while the short durations are reminiscent of the black hole transient
V4641 Sgr.

A fraction of the new sources has been detected in earlier
observations with instruments on ASCA, BeppoSAX and RXTE. We here
discuss the earlier detections with the BeppoSAX Wide Field
Cameras. Thanks to its large field of view (40$^{\rm o}\times40^{\rm
o}$ with 5$^\prime$ angular resolution, similar to IBIS) and a
bandpass that extends to the unobscured $>10$~keV part, the
contribution of the WFCs to the understanding of these transients may
be relevant. For instance, an accurate flux history of these hard
sources will provide strong constraints on the suggested transient
nature of some sources.

The WFCs were operational between June 1996 and April 2002, in other
words in the six years preceding the launch of INTEGRAL.  The WFC data
archive has a substantial coverage over large parts of the sky. Along
the Galactic plane, the exposure time is always in excess of 1 million
seconds. The peak of the coverage coincides with the Galactic center
where 6 million seconds of net exposure time was obtained. Typical
sensitivities are a few mCrab for months-long observation times. For a
recent review of these observations, we refer to In 't Zand et
al. (2004).

\begin{table*}[!t]
  \begin{center}
    \caption{List of transients.\label{tab1}}\vspace{1em}
    \renewcommand{\arraystretch}{1.2}
    \begin{tabular}[h]{llccll}
      \hline\hline
      IGR Name & Alternative name  & Follow & WFC de- & WFC average  & Remark \\
and ref.          &                   & up$^\ddag$     & tection?& flux$^\dag$  &        \\
      \hline
J06074+2205$^1$ &                           &   &   & $<9$ (3.2) & \\                 
J15479-4529$^2$ & 1RXS J154814.5-452845     &   &   & $<1.5$ (2.1) & Interm. polar \\   
J16316-4028$^3$ &			    &   &   & $<1.2$ (4.0) & =3EG 1631-4033? \\ 
J16318-4848$^4$ &                 	    & X &   & $<0.7$ (1.9) & Strong obscuration \\  
J16320-4751$^5$ & AX J1631.9-4752	    &R,X& y & $3.3\pm0.2$ (1.9) & \\                 
J16358-4726$^6$ &			    &C,X&   & $<0.5$ (2.1) & 5.5 ksec oscillation \\
J16393-4643$^7$ &			    &   &   & $<0.9$ (2.0) & =3EG 1639-4702? \\ 
J16418-4532$^2$ &			    &   &   & $<2$ (4.2) & \\                 
J16479-4514$^8$ &			    &   &   & $<2$ (4.4) & Transient \\                 
J17091-3624$^9$ & 1SAX J1709-36	            &   & y & $<2.4$ (4.4) & Transient \\                 
J17391-3021$^{10}$ & XTE J1739-302	    &   &   & $<1.3$ (4.0) & Fast transient \\  
J17456-2901$^{11}$ &			    &   &   & $<1.3$ (4.0) & Gal. Ctr. source \\   
J17464-3213$^{12}$ & H1743-322		    & C,R & & $<1.6$ (4.0) & BH transient \\              
J17475-2822$^{13}$ &			    &   &   & $<1.7$ (4.0) & \\                 
J17544-2619$^{14}$ &			    & X & y & $<0.5$ (3.5) & Fast transient \\  
J17597-2201$^{15}$ & XTE J1759-220	    & R &   & $<2.0$ (3.3) & Burster, 1-3 hr orbit \\ 
J18027-2017$^{13}$ & SAX J1802.7-2017	    &   &   & $<17$  (3.3) & Pulsar in HMXB \\ 
                   &                        &   &   &              & close to GX 9+1 \\                 
J18325-0756$^{16}$ &			    &   &   & $<1.4$ (1.1) & Transient \\                 
J18483-0311$^{17}$ & 1RXH J184817.3-031017  &   & y & $2.2\pm0.3$ (0.9) & Transient \\                 
J18539+0727$^{18}$ &			    &   &   & $<0.7$ (1.5)	& Transient \\                 
J19140+098$^{19}$  & EXO 1912+097           & R & y & $1.0\pm0.1$ (1.5) & \\                 
      \hline\hline \\
      \end{tabular}
    \label{tab:table}
  \end{center}

\noindent
$^1$Chevenez et al. 2004; $^2$Tomsick et al. 2004; $^3$Rodriguez et
al. 2003a; $^4$Courvoisier et al. 2003; $^5$Tomsick et al. 2003;
$^6$Revnivtsev et al. 2003c; $^7$Malizia et al. 2004; $^8$Molkov et
al. 2003; $^9$Kuulkers et al. 2003; $^{10}$ Sunyaev et al. 2003a;
$^{11}$B\'{e}langer et al. 2004; $^{12}$Revnivtsev et al. 2003a;
$^{13}$Revnivtsev et al. 2004; $^{14}$Sunyaev et al. 2003b;
$^{15}$Lutovinov et al. 2003a; $^{16}$Lutovinov et al. 2003b;
$^{17}$Chernyakova et al. 2003; $^{18}$Lutovinov et al. 2003c;
$^{19}$Hannikainen et al. 2003, Cabanac et al. 2004a,b and Schulz et al. 
2004

\noindent
$^\dag$These are the 6-year average fluxes (mCrab in 2-28
keV). Between parentheses are indicated the exposure times checked, in
Msec. The sensitivity also depends on the off-axis angle.

\noindent
$^\ddag$X-ray follow-up activity up to Feb. 1, 2004 (X=XMM-Newton, R=RXTE,
C=Chandra)
\end{table*}

\section{WFC analysis}

We have gone through all WFC data of the 21 IGR sources and searched for
detections on time scales of 1.5~hr, 1~d, 6 months and 6 years and in
energy bands 2-10 and 10-25 keV. The total exposure time checked
ranges from 1 to 4 million seconds. Per source this is up to 30\% less
exposure than present in the archive. This is due to conservative data
filter thresholds, to avoid any complication due to partly obscuration
by the earth of the field of view. Table~\ref{tab1} provides the
average fluxes of all 21 sources. We found detections in 5 cases.  In
all other cases the upper limits over all observations combined are
between 0.7 and 2.4 mCrab (2-28 keV).

In the following we discuss some results on the five detected
sources. In particular, we discuss the 2-28 keV flux history.
Beforehand we note that none of these 5 sources exhibited type-I X-ray
bursts.

\subsection{IGR J16320-4751}

IGR J16320-4751 = AX J1631.9-4752 (Tomsick et al. 2003) was
consistently detected with the WFCs during 1996-2002, at a flux that
varied between 10 and 16 mCrab (5-10 keV; 2-month averages); with the
ASCA detection, this suggests that the source has been persistently
active for at least 8 years.  Based on three broad channels (2-5,
5-10, and 10-28 keV) and averaged over all data, an absorbed power law
model describes the data well with a photon index of 2.5 $\pm$ 0.3 and
$N_{\rm H} = (2 \pm 1) \times 10^{23}$ cm$^{-2}$.  The average
unabsorbed 0.7-10 keV flux is $4 \times 10^{-10}$ \ecs.  If the source
is indeed a high-mass X-ray binary (HMXB), as suggested by Rodriguez
et al. (2003b), the persistently bright emission would be more in line
with the mass donor being a (super)giant than a Be star.  The position
in the combined WFC data is $\alpha_{2000.0} = 16^{\rm h}32^{\rm
m}05\fs4, \delta_{2000.0} = -47^{\rm o}52'07"$ (error radius 1\farcm7;
all WFC positional accuracies are given at 99-percent-confidence),
which is 1\farcm1 from AX J1631.9-4752 (ref TBD) and 1' from IGR
J16320-4751.  The analysis of an XMM spectrum by Rodriguez et
al. (2003b) is in general agreement with the WFC measurements, but
ASCA measurements show a much harder spectrum with an index of 0.5.

\subsection{IGR J17091-3624}

During Sept. 20.7--21.7 and Sept. 29.8 through Oct. 1.1, 2001 (UT),
the WFCs detected a source which coincides with IGR J17091-3624. From
combined data taken during August-October 2001, the source was
localized to $\alpha_{2000.0} = 17^{\rm h} 09^{\rm m} 06^{\rm s},
\delta_{2000.0} = -36^{\rm o}24'39"$, with an error radius of
1\farcm5. The position is 0.03' from the INTEGRAL centroid (Kuulkers
et al. 2003), 0.5' from the COMIS-TTM centroid (Revnivtsev et
al. 2003b), and 1.3' from the reported radio source (Rupen et
al. 2003). The flux was 14 mCrab and 20 mCrab (2-10 keV) during the
two observations. The spectrum is consistent with a 3.0$\pm$0.4
photon-index power law or k$T$=4.3$\pm$1.4 keV for thermal
bremsstrahlung; $N_{\rm H}$ was constrained to upper limits of 5 and
$2\times10^{22}$~cm$^{-2}$, respectively (90\% confidence). A search
through the WFC archive revealed a weak detection five years earlier,
in the combined data of the 1996 August-Oct WFC campaign of the
Galactic center. In 650 ksec of data, the source was detected at an
average flux of about 5 mCrab. All X-ray data combined imply that IGR
J17091-3624 is a moderately bright variable X-ray source which flared
in Oct. 1994 (Mir-COMIS/TTM; Revnivtsev et al. 2003b), Sept. 1996
(BeppoSAX-WFC), Sep. 2001 (BeppoSAX-WFC), and April 2003
(INTEGRAL-IBIS) and most likely is an X-ray binary. The spectrum is
fairly soft and no X-ray bursts were detected which marginally hints
at a black hole transient at a relatively far distance.  We note that
the source was followed up by the BeppoSAX Narrow-Field Instruments on
Sept. 25, 2001, after the first real-time detection. Therefore, it is
listed in the SAX observation catalog under 1SAX
J1709-36. Unfortunately, this observation was terminated before useful
data could be taken due to problems with the satellite attitude
control.

\subsection{IGR J17544-2619}

\begin{figure}[!t]
\centering
\includegraphics[height=0.99\columnwidth,angle=270]{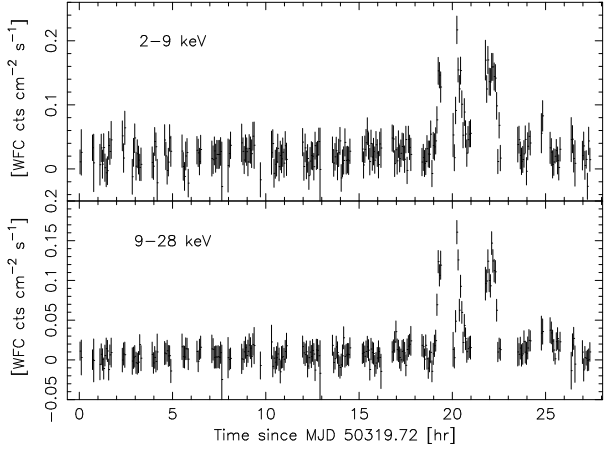}
\caption{Long flare from IGR J17544-2619 on August 24,
1996. The time resolution is 4~min.
\label{fig1754lc1}}
\end{figure}

\begin{figure}[!t]
\centering
\includegraphics[height=0.99\columnwidth,angle=270]{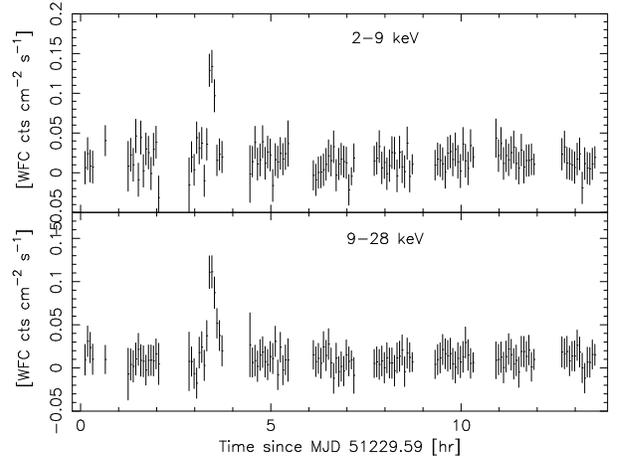}
\caption{Short flare from IGR J17544-2619 on Feb. 20,
1999. \label{fig1754lc2}}
\end{figure}

\begin{figure}[!t]
\centering
\includegraphics[width=0.99\columnwidth]{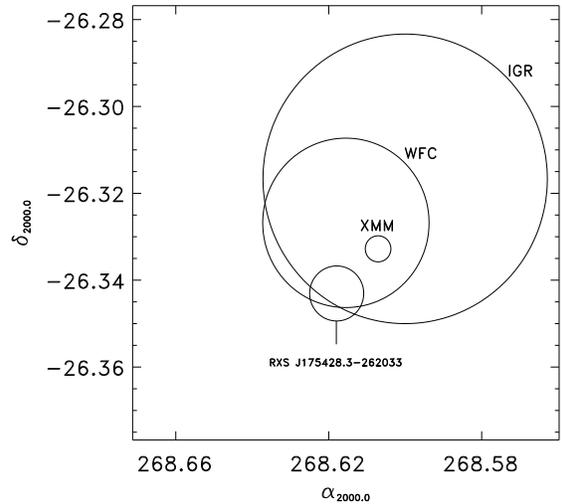}
\caption{Map showing source detections around J17544-2619. All error
circles are for a confidence level in excess of 90\%.\label{k1754}}
\end{figure}

\bron\ is a flaring source which was first detected with INTEGRAL on
Sep. 17, 2003, when it exhibited two flares (Sunyaev et al. 2003b
Grebenev et al. 2003) and a second time on Mar. 8, 2004 (Grebenev et
al. 2004). The first detection was accompanied by two XMM-Newton
observations which revealed further strong, flaring-like, activity at
flux levels one to two orders of magnitude fainter (Gonzalez-Riestra
et al. 2003, 2004). The nature of the source is thus far
undetermined. However, there is the suggestion from the strong
variability and the small angular distance to the Galactic center
(3\fdg3) that it is a compact object in our Galaxy.

Actually, \bron\ was detected by the WFCs already in August 1996. The
detection lasted just a few hours with a peak flux of about 50~mCrab
(Fig.~\ref{fig1754lc1}). The combination of duration and peak flux
prevented real-time searches from finding the source and it was only
noticed during more sensitive off-line searches weeks later. By that
time it was obviously too late for a target of opportunity
observation. The source was tentatively identified with a ROSAT source
detected during the all-sky survey, 1RXS J175428.3-262033, as was also
done after the first INTEGRAL detection (Sunyaev et al. 2003b) by
Wijnands (2003). The source was detected a further 4 times and each
time it was only noticed weeks to months later during offline
searches. Due to its presumed association with the ROSAT source (see
Fig.~\ref{k1754}) and in anticipation of a number of pending follow-up
requests it was not reported.  Only due to the INTEGRAL detection and
the subsequent XMM-Newton observation it was established that this
source is actually different from the ROSAT source (Gonzalez-Riestra
et al. 2003, 2004).

Table~\ref{tab2} shows the characteristics of the 5 flares observed
from IGR J17544-2619 with the WFCs. Figs.~\ref{fig1754lc1} and
\ref{fig1754lc2} show the light curves of the the shortest and longest
flare. The durations vary widely between 10~min and 8~hr. The peak
flux was measured to be between 100 and 200 mCrab (2-28 keV). The
flares appear clustered in time. Flares 2/3 and 4/5 are within 20~d
from each other.

\begin{figure}[!t]
\centering
\includegraphics[width=0.99\columnwidth]{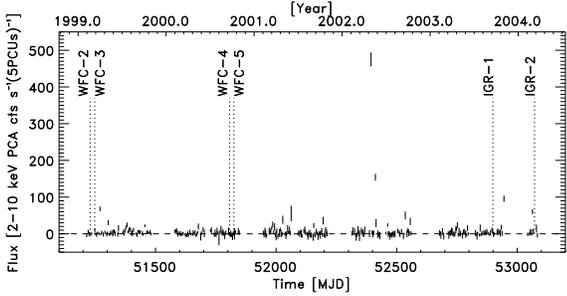}
\caption{Light curve of IGR J17544-2619 from RXTE-PCA bulge scans.
The times of bright flares (i.e., brighter than 0.1 Crab) seen with
BeppoSAX and INTEGRAL are indicated. 10.4~c~s$^{-1}$5PCU$^{-1}$ is
equivalent to 1 mCrab.\label{bulge}}
\end{figure}

\begin{figure}[!t]
\centering
\includegraphics[width=0.99\columnwidth]{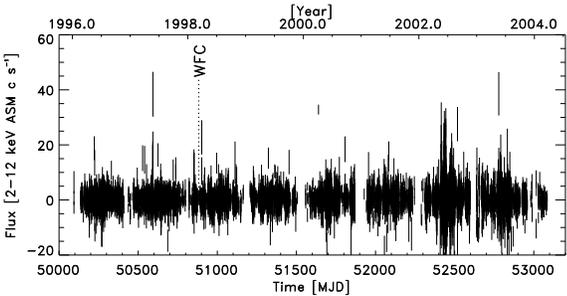}
\caption{Light curve of SAX J1818.6-1703 from RXTE-ASM 90-s dwells.
The times of the one bright flare seen with BeppoSAX is
indicated. 7.5~c~s$^{-1}$ is equivalent to 100 mCrab.\label{asm1818}}
\end{figure}

The spectrum of flares 3 and 4 cannot be modeled by bremsstrahlung or
a power law. It is fairly well modeled by absorbed black body
radiation with $N_{\rm H}$ fixed to $1.3\times10^{22}$
(Gonzalez-Riestra et al. 2003, 2004) and k$T=3.4\pm0.1$~keV with
unabsorbed peak 2-28 fluxes of 5.4 and 7.2$\times10^{-9}$~\ecs. When
extrapolated to the 18-25 keV band, the flux is 228 mCrab and 369
mCrab (53 and 93 mCrab in 25-50 keV). These numbers are comparable
with the numbers for the two flares seen with INTEGRAL in Sep. 2003
(Sunyaev et al. 2003a; Grebenev et al. 2003), certainly if those
numbers represent fluxes averaged over the flares. The black body
model is not the only good model. An absorbed power law with a
high-energy cutoff fits just as well and the extrapolated fluxes in
the two INTEGRAL bands do not differ by more than 30\%.

There are several data sets that support the notion that even outside
large flares, the flux also appears to vary violently. There are the
two XMM-Newton observations on Sep. 11 and 17, 2003, when for instance
the flux increased from 8.8$\times10^{-13}$ to
4.0$\times10^{-11}$~\ecs\ (0.5-10 keV) in a matter of $\sim$10 minutes
(Gonzalez-Riestra et al. 2004). Furthermore, the source was not
detected during a serendipitous XMM-Newton observation in March 2003
with an upper limit of 5$\times10^{-14}$~\ecs. Then there are the
semi-weekly flux measurements through scans with the RXTE Proportional
Counter Array of the Galactic bulge (Swank \& Markwardt 2001), see
Fig.~\ref{bulge}. During the 5 years that these measurements have
persisted so far, \bron\ flared up above the 1.5~mCrab detection
threshold (about 10$^{-11}$~\ecs\ in 2-10 keV) 16 non-consecutive
times.  This points to a duty cycle above that level of about
5\%. Finally, there are measurements with the RXTE All-Sky Monitor,
which occasionally show fluxes up to 0.4 Crab (2-12 keV) during 90-s
dwells.

In conclusion, \bron\ is a source which appears to be active for a
large portion of the time with strong variability between an upper
limit of 5$\times10^{-14}$~\ecs\ and a peak flux of
$\sim10^{-9}$~\ecs. No X-ray bursts were ever detected.  This kind of
behavior is quite reminiscent of
V4641~Sgr=SAX~J1819.3-2525=XTE~J1819-254 (Wijnands \& van der Klis
2000, In 't Zand et al. 2000, Hjellming et al. 2000) which is a
dynamically established black hole candidate (Orosz et al. 2001). In
that object also strong variability was detected with hours-long flare
activity that culminated in a one of the brightest X-ray flares ever
observed (12 Crab units). Therefore, perhaps this is a black hole
candidate, as was also recently suggested by Gonzalez-Riestra et
al. (2004). The variability would then be due to an atypical companion
star. Optical follow-up may bring the solution. A fairly bright O-type
candidate has been identified by Rodriguez (2003) and Gonzalez-Riestra
et al. (2004). A search for orbital Doppler shifts will likely provide
constraints on the mass of the compact object.

We note that the list of similar fast X-ray transients without obvious
optical counterparts such as RS CVn stars, BY Dra flare stars, or
pre-main sequence objects, is slowly growing. To our knowledge, it now
consists of five objects: IGR J17544-2619, XTE J1739-302 (Smith et
al. 1998), V4641 Sgr (In 't Zand et al. 2000; Orosz et al. 2001), XTE
J1901+014 (Remillard \& Smith 2002), and SAX J1818.6-1703 (In 't Zand
et al. 1998). The latter source was seen only once with the WFCs, but
an investigation of the RXTE/ASM light curve (Fig.~\ref{asm1818})
shows at least 3 more flares above 100 mCrab (2-12 keV). Therefore,
all these sources exhibit multiple hours-long flares with peak
fluxes around $10^{-9}$~\ecs. For the first three systems there is now
evidence that the strong variability may be due to a high-mass
companion star which may feed the compact object for a significant
part through a wind instead of an accretion disk (like in many
high-mass X-ray binaries).  None of the systems exhibited X-ray
bursts. When taking into account that the WFC exposure of these
sources is a few million seconds, this provides further evidence that
the compact objects are black hole candidates. There is a subtle
non-uniformity in this group: V4641~Sgr, the only dynamically
confirmed black hole candidate, is not {\em continuously} active with
a small duty cycle, but {\em intermittently} with a high duty
cycle. The key to understanding these systems is optical follow up
(see above).

\begin{table}
\caption[]{Flares from IGR J17544-2619.\label{tab2}}
\begin{center}
\begin{tabular}{ccc}
\hline\hline
Time & Peak flux & Duration$^\ddag$ \\
(MJD) & (WFC c~s$^{-1}$cm$^{-2}$) & (hr) \\
\hline
50320.6 & 0.38 & $3.3\pm0.2$ \\
51229.7 & 0.31 & $0.2\pm0.2$  \\
51248.8 & 0.20 & $0.4\pm0.2$ \\
51807.5 & 0.25 & $8.3\pm0.2$ \\
51825.1 & 0.38 & $1.0\pm0.2$ \\
\hline\hline
\end{tabular}
\end{center}

\noindent
$^\ddag$time when flux is in excess of 0.1 times
the peak value.
\end{table}

\begin{figure}[!t]
\centering
\includegraphics[width=1.\columnwidth]{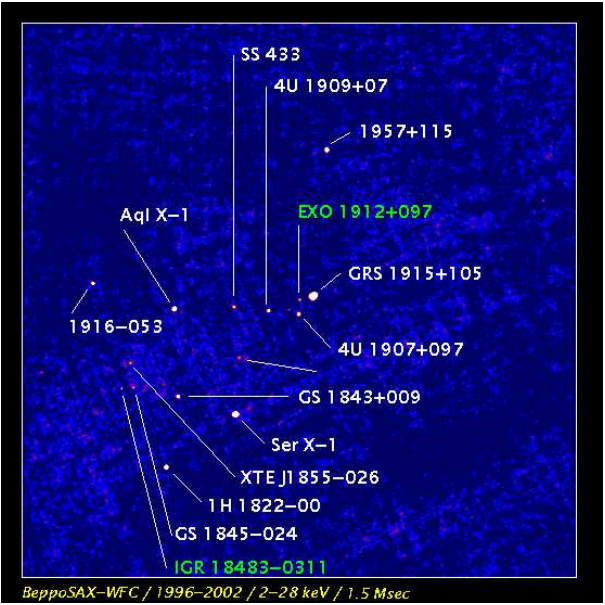}
\caption{BeppoSAX-WFC combined image for 1996-2002 data with
IGR~J18483-0311 and EXO 1912+097 contained.
\label{ssum2}}
\end{figure}

\begin{figure}[!t]
\centering
\includegraphics[width=0.99\columnwidth]{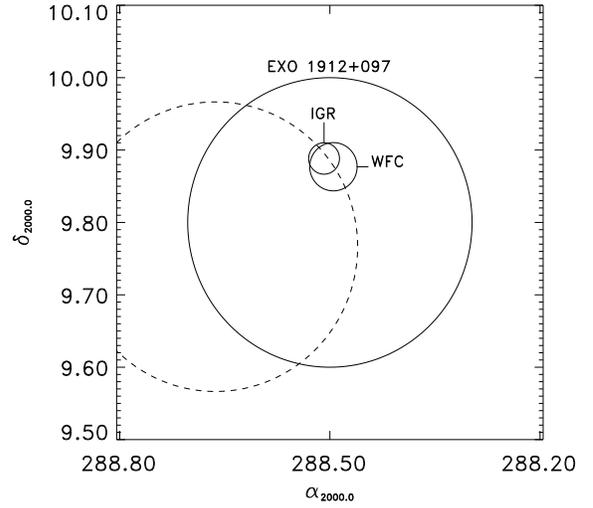}
\caption{Map showing source detections around J19149+098.  There are
two positions for EXO 1912+097. The dashed circle refers to the
coordinates as provided by Simbad, the solid (large) circle to those
inferred from Lu et al. (1997; this is a paper written in
Chinese). The INTEGRAL data (JEM-X and IBIS combined) are from Cabanac et
al. (2004b).\label{k1914}}
\end{figure}

\subsection{IGR J18483-0311}

The coverage of BeppoSAX-WFC on IGR J18483-0311 is limited
due to frequent large off-axis angles.  The maximum flux observed with
BeppoSAX-WFC is 0.03 c~s$^{-1}$cm$^{-2}$ (2-28 keV), or 15 mCrab, on
April 26, 1997, and the lowest flux 1.5 mCrab for the combined data of
Jul-Dec 1996 (257 ks exposure). All upper limits are worse than the
lowest detected flux. The 6-year average is $2.2\pm0.3$ mCrab. Since
it is located in the Scutum arm of our Galaxy where many HMXBs were
detected in the past (cf Koyama et al. 1990; see Fig.~\ref{ssum2}), it
may also be a HMXB.

\subsection{IGR J19140+098}

IGR J19140+098 was detected in IBIS in March 2003 (Hannikainen et
al. 2003) and May 2003 (Cabanac et al. 2004a). The peak flux was 0.1
Crab (13-100 keV). A brief (3 ksec) follow-up observation was
performed by RXTE in March 2003 when it peaked at about 10 mCrab in
2-10 keV (Swank \& Markwardt 2003). The spectrum is variable
with a transient Fe-K emission line complex (with equivalent width 550
eV during the RXTE observation) and a power law with photon index
between 1.7 and 3. The source shows significant variability on time
scales longer than 100~s.

This source also has limited coverage by the BeppoSAX WFCs. All 1-d
WFC detections cluster around 10 mCrab (similar to what was observed
with RXTE), and the 6-month averages vary between an upper limit of 1
mCrab (3$\sigma$) and a detection of $5.0\pm0.5$ mCrab (for Jan-Jun
1998). The position is $\alpha_{2000.0} = 19^{\rm h}13^{\rm m}09\fs5,
\delta_{2000.0} = +9^{\rm o}52'37"$ (error radius 2'). The 6-year
average is $1.0\pm0.1$ mCrab (exposure time 1.5 Msec; see
Fig.~\ref{ssum2}).  All WFC detections are insufficient to provide
meaningful constraints on pulse signal amplitudes or spectra.

Based on coordinates extracted from the Simbad database at CDS
Strasbourg, we identified this source with a EXO~1912+097. This is a
source which was identified in EXOSAT data in the nineties through a
new detection algorithm by Lu et al. (1997). However, investigation of
the Lu et al. paper, which is written in Chinese, yielded that the
error circle of EXO 1912+097 is actually shifted by such an amount
that the WFC and INTEGRAL source positions are on the edge of the
error circle. Therefore, it is slightly questionable whether the
EXOSAT source is related to the WFC and INTEGRAL sources. For details,
see Fig.~\ref{k1914}.

The source was often detected with WFC and INTEGRAL, but at flux
levels close to the detection limit. Therefore, it is feasible that
the source is strongly variable but persistent. A determination of its
physical nature would benefit from high-spatial-resolution
observations at X-ray and optical wavelengths, also to confirm the
tentative identification with a radio counterpart (Schulz et al. 2004;
Pandey et al. 2004).

\section{Summary}

BeppoSAX-WFC provided detections of 5 out of 21 of the new INTEGRAL
sources in the classical X-ray band. The flux levels of these
detections are all except one below $10^{-10}$~\ecs. The one exception
is a clear transient which goes into outburst for a few hours at least
once a year. One other source shows the hallmarks of a classical
though faint transient: IGR J17091-3624. The other three sources
exhibit detections in the combined 6-year WFC data which, for fluxes
which were rarely measured to be above a factor of 10 of the 6-year
averages, implies that they are most likely low-flux persistent
sources.
The two transients are very likely X-ray binaries in our galaxy. The
nature of the three fainter but possibly persistent sources is less
well constrained, but probably also X-ray binaries in the Galaxy.

\end{document}